# A remote reactor monitoring with plastic scintillation detector

A. Sh. Georgadze, V. M. Pavlovych, O. A. Ponkratenko, D. A. Litvinov

*Abstract*— Conceiving the possibility of using plastic scintillator bars as robust detectors for antineutrino detection for the remote reactor monitoring and nuclear safeguard application we study expected basic performance by Monte Carlo simulation. We present preliminary results for a 1 m$^3$ highly segmented detector made of 100 rectangular scintillation bars forming an array which is sandwiched at both sides by the continuous light guides enabling light sharing between all photo detectors. Light detection efficiency is calculated for several light collection configurations, considering different scintillation block geometries and number of photo-detectors. The photo-detectors signals are forming the specific hit pattern, which is characterizing the impinging particle. The statistical analysis of hit patterns allows effectively select antineutrino events and rejects backgrounds. To evaluate detector sensitivity to fuel isotopic composition evolution during fuel burning cycle we have calculated antineutrino spectra. The statistical analyses of the antineutrino spectra for several dates of fuel burning cycle, folded with 15% FWHM at 1 MeV energy resolution of the detector, prove the possibility of proposed detector to measure fuel composition evolution during fuel cycle.

## I. INTRODUCTION

Nuclear reactors are the most intense artificial antineutrino sources emitting around 10$^{20}$ antineutrinos per second for a 2-3 GW$_{th}$ power plant. In a water reactors, the main contribution (>more then 99%) to the power production is caused by the fissioning of the four main fuel isotopes, $^{235}$U, $^{238}$U, $^{239}$Pu, and $^{241}$Pu. An average energy of about 200 MeV released per fission and about 6 neutrinos produced along the β-decay chain of the fission products.

Since antineutrinos weakly interact with matter, the antineutrino flux is not affected by the core neutron-gamma radiation shielding and therefore the total antineutrino flux allows the independent on-line monitoring of the integrated reactor thermal power. In addition there is a possibility to retrieve the core isotopic composition from the antineutrino energy spectrum analysis. During a reactor cycle due to burning-up of $^{235}$U and accumulation in the active area of reactor of $^{239}$Pu, the spectrum of antineutrinos from reactor changes with a time. This is because the fission products of $^{235}$U and $^{239}$Pu have different beta-decaying neutron rich fission fragments, therefore β-decays result in different antineutrino energy spectra.

The concept of antineutrino power reactor monitoring was first proposed by Mikaelyan and pioneering experiments were performed at the Rivne Nuclear Power Plant in Ukraine [1]. The power measured with antineutrinos has agreed with the thermal measurements within 2.5% and the effect due to changing uranium and plutonium content was demonstrated [2]. Recently, the practical feasibility of reactor monitoring using antineutrinos has also been demonstrated using a tone-size detector at the San Onofre power station, called SONGS [3].

Sh. Georgadze, V. M. Pavlovych, O. A. Ponkratenko are with Institute for Nuclear Research, National Academy of Sciences of Ukraine, Kyiv (e-mail: georgadze@kinr.kiev.ua; pavlovich@kinr.kiev.ua; ponkrat@kinr.kiev.ua).

D. A. Litvinov is with Institute for Safety Problems of Nuclear Power Plants, National Academy of Sciences of Ukraine, Kyiv (e-mail: litvinovda@mail.ru)

In these experiments the liquid scintillator detectors have been used for antineutrino reactor monitoring. But liquid scintillators are toxic and flammable therefore their use on commercial power plants is restricted.

As alternative, plastic scintillators as a hydrogen rich target were proposed [4,5] to solve detector safety issues.

When consider antineutrino detector as a regular tool for nuclear reactor monitoring, the requirements of simplicity design and operation, reliability during years of continuous operation, moderate costs are appears.

Starting from these requirements we have developed highly segmented antineutrino detector, based on the use of plastic scintillator in form of prismatic bars packed to an array and sandwiched on both sides by rectangular continuous light guides which enables the light-sharing between photo-detectors. Such a solution for detector makes possible to use less photo-detectors comparing to case when two photo-detectors attached to each scintillation bar. This feature of the proposed detector is based on the concept of block-design scheme widely used in Positron Emission Tomography scanners. Such a design is satisfying of terms of the simplicity design and is essentially cost effective, otherwise enabling more stable detector operation due to decreasing number of crucial parts.

As a feasibility study for setting up an new experiment at the Rivne Nuclear Power Plant (NPP) for sterile neutrino search and reactor monitoring we have study the expected basic performance of the proposed detector and have optimize the detector geometry and parameters by Monte Carlo simulation.

## II. MATERIALS AND METHODS

### A. Description of the Detector

The detector shown in Fig. 1 consists of 100 rectangular plastic scintillator bars of 10 cm ×10 cm × 100 cm size packed to cubic block of 1 m$^3$ active volume and sandwiched on both sides by rectangular continuous light guides that enable light-sharing between photo-detectors. The number of PMTs with diameter 200 mm (or 130 mm) is 16 on each side, with a total amount of 32. To improve light collection, the photo-detectors are coupled to light guides through light concentrators. The plastic scintillator bars have smooth cut surfaces on all sides with aluminized Mylar films and gadolinium (Gd) coated Mylar films on each side except for the one coupled to the light guide. To reduce the cosmogenic and reactor induced backgrounds, the detector is surrounded by passive shield from lead and borated polyethylene. The continuous rectangular light guides are shielding the detector from γ-quanta, emitted by radioactive impurities, present in PMTs. Cosmic rays can be discriminated by software analyzes because of the large deposited energy (a muons crossing the plastic scintillator releases ~2 MeV/cm) and event topology (a charge particle passing through the detector would create a hit in each bar along the path).

Antineutrinos from a reactor are detected via (IBD) reaction:

$$\widetilde{\nu}_e + p \rightarrow e^+ + n \qquad (1)$$

with an energy threshold of 1.8 MeV and subsequent reactions:

$$e^+ + e^- \to 2\gamma,$$
$$n + {}^{155}Gd \to {}^{156}Gd^* \to {}^{156}Gd + \gamma,$$
$$n + {}^{157}Gd \to {}^{158}Gd^* \to {}^{158}Gd + \gamma.$$

The reaction threshold is 1,806 MeV. The antineutrino $\nu_e$ interact with proton-rich scintillator producing a positron and a neutron. Positron, slowing down in scintillator due to ionization processes fully releases its energy and almost at rest annihilates with an electron with the radiation in opposite directions of two gammas of 511 keV each.

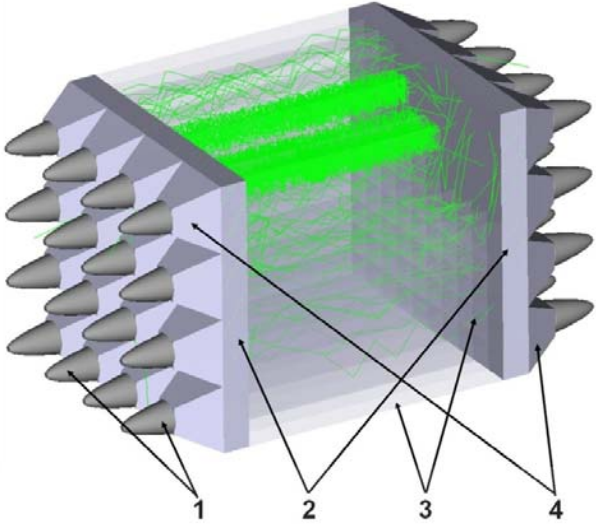

**Figure 1.** A conceptual design of the antineutrino detector: 1 – PMTs; 2 – flat light guides; 3 – plastic scintillation bars; 4 – light concentrators.

The IBD signals can be discriminated from cosmogenic and reactor induced backgrounds by standard method of delayed coincidence of prompt positron event accompanied with annihilation gammas, depositing by ionization process and detected in the energy window between 2−8 MeV and delayed neutron capture on Gadolinium event, depositing in the energy window between 3−8 MeV in a time window of ~ 4-200 μs.

Due to high detector segmentation the prompt and delayed events are correlated both temporally and spatially, providing a tag for IBD events selection. External and internal gammas and neutrons produce multiple events in scintillation bars, which also allow topological background cuts based on segmentation.

In the proposed antineutrino detector design, due to the light sharing qualities of the continuous light guides implemented in, the optical photons from events in scintillation bars are detected by all PMT simultaneously. Signal amplitude of a PMT depends on the location of points of interaction of secondary particles from events and resulted in distribution of PMT signals of a 16×16 PMT array. The light intensity ratio seen by group of PMTs on each side of a scintillation bar array allows one to resolve the 'center of gravity' of several interactions in the scintillation bars array and summing all photo-detector signals to measure energy of event.

The developed detector scheme is unique and differs from methods in-use for other antineutrino detectors. In particular, in PANDA detector [4] the scintillation signal in every individual bar is detected by two PMTs settled on both ends that resulted in large number of output channels and high detector cost. In other case, the project of CoRMORAD [5], several bars are assembled to cluster, which is viewed on both sides by PMT. In this case there is unjustified loss of segmentation that resulted in loose of additional criteria of topological selection of antineutrino events from the background.

Compared to the both ends readout on each scintillator bar detector design applied in PANDA detector, the implemented "block detector" design in proposed detector allow the same power of event selection, but results in significantly fewer PMTs, 32 instead of 200, simplification of data acquisition system and therefore reduced detector cost.

Such a detector design is referred to a pixelated block detectors that is widely used in for medical imaging in positron emission tomography [6] is resulted in a drastic decrease in number of output readouts comparing to both ends coupling between scintillation bar and two photo-detectors that makes the data acquisition system much cheaper.

*B. Detector simulation and optimization.*

Full simulations of the detectors were performed using both MCNP particle transport code [7]. The ZEMAX [8] software was used to simulate light transport in the detector. The simulations were intended to optimize the relevant parameters of the layout such as the bar size, the number of bars, the number of PMT, light collection efficiency.

Generated particles, as well as secondary produced by interaction in the detector, were tracked step-by-step using a MCNP particle transport code to determine the energy deposited in the plastic scintillator and to fix points of interactions of particles with detector material. At each step, the number of optical photons emitted by the active material was evaluated and transferred to ZEMAX light transport code to calculate light collection and estimate signal detected by the 16 × 16 PMT array, using effective parameters to account for the light attenuation in the scintillator and quantum efficiency of PMTs.

We have built in ZEMAX environment the detailed optical model of the detector to determine the light collection efficiency and to optimize geometry of detector modules. We investigate the effect of PMT number and its size, rectangular cross section of scintillation bars and varied light guide thickness on the scintillator bars identification accuracy and position resolution of a scintillation bars array.

The simulation were performed for two PMT sizes of 5" and 8" inches and for three different light guide thicknesses for each of the PMT size, and additionally thickness of continuous light guide varied for 5 cm, 10 and 15 cm.

The scintillation bars supposed to be made of PS-923A [9] plastic scintillator produced by Amcrys company (Ukraine) were modeled as smooth with specular reflections, but adding diffuse (Lambertian) reflections to account non ideal surface treatment and wrapped first in aluminized Mylar film, separated from the scintillator by an air gap, with reflection coefficient $RC = 0.85$ and then covered with gadolinium coated Mylar film considered in the model as ideal absorber. The data on bulk attenuation length (BAL) of the scintillation 250-450 cm and light output 59% of anthracene, that correspond to $\sim 10^4$ optical photons/MeV were taken from [9]. A concentrators and continuous light guides, enabling light sharing, were simulated with data supplied by ZEMAX glass library. The PMTs were modeled as a 1 mm thick disk of varying size with 100% active area, made of borosilicate glass BK7 with refractive index n = 1.52. The light detector was placed on outer surface of the PMT glass BK7 and assigned as ideal absorber with 100% active surface and 100% efficiency to detect optical photons.

The simulated light collection coefficients for 8 inch 16 ×

16 PMT array and three scintillation bars cross-sectional dimensions of 5 cm × 5 cm, 7 cm × 7 cm and 10 cm × 10 cm are 0.10, 0.15 and 0.19 correspondingly and for 5 inch 16 × 16 PMT array the light collection coefficients are 0.05, 0.08 and 0.11.

## III. RESULTS AND DISCUSSION

### A. Signal selection and background reduction

*1) Detector geometry:* We select the 10 cm × 10 cm cross-sectional dimension of rectangular square bar type scintillation module which offer reasonable handling and detector cost. With scintillation bar cross section of 10 × 10 cm and number of PMT 16 on each side of detector the resulted light collection efficiency based on ZIMAX simulation of ≈ 20% is high enough for good energy resolution and appropriate mean time of neutron capture of about 60 μs [4].

*2) IBD signal selection:* There are two main background sources which can mimic antineutrino signatures: the accidental background and the correlated background. The correlated backgrounds are induced by cosmic muons and are consisting of two groups: the background due to β-n decaying cosmogenic isotopes and fast neutrons. The uncorrelated background is caused by accidental coincidences of prompt event produced by natural radioactivity from of construction materials and delayed event produced by captured neutrons produced mainly by cosmic ray muons

For prompt event the energy of the positron from IBD reaction during ionization, Bhabha scattering, and bremsstrahlung processes is absorbed in the plastic scintillator bar, then the slowing down positron subsequently annihilates producing two gammas of 511 keV each with opposite directions. The 511 keV gammas mainly interact via Compton scattering with the plastic scintillator producing several electrons, which are absorbed in a mm range. Thus, positron together with 511 keV gammas produces several flashes in neighboring scintillation bars that are detected by PMTs. The antineutrino's energy $E$ is measured from the total energy deposited in the detector. In ZEMAX for these flashes were attributed the spherical light sources with intensity in accordance to energy released.

The simulated in ZEMAX transport of rays attributed to scintillation light fleshes caused by all primary and secondary particles is resulted in amount of rays detected by each photo-detector. After applying 25% PMT light detection efficiency a 16 × 16 array PMT signals distribution in units of photoelectron number can be built, forming a specific hit pattern, which is a characteristic of incident particle.

The IBD prompt event topology is formed by strong signal from positron and weak signals of Compton electrons from two 511 keV annihilation gammas that are interacted in the detector in spatially adjacent scintillation bars. Thus the antineutrino event selection is required to have the specific topology of interest, which is characterized by specific hit pattern. The analysis of event hit pattern can be used to extract event type information and reject background events.

To obtain volumetric response of the detector to IBD reaction the prompt events were simulated by moving positron interaction point along the scintillation bars with 5 mm step. Detector energy response was obtained for prompt events in energy range of 2 – 8 MeV with the step of 1 MeV but can be interpolated for arbitrary energy. The total numbers of $2 \cdot 10^3$ prompt events were simulated to evaluate detector respond to IBD reaction. The simulated data create dataset which is used to perform selection cuts by comparing with the help of statistical algorithm the hit pattern of detected signal with simulated one, expected from prompt event of IBD reaction.

From Fig. 2 one can see that hit pattern of prompt IBD event with energy of 8 MeV is sharp and peaked, unlike to wide hit pattern of Fig. 3 of delayed event formed by 8 MeV gamma ray cascade caused by neutron capture on Gd. Performing statistical analysis of PMT signals distributions (hit patterns) for prompt and delayed events give us a quantitative measure of topology difference of electronic signals produced by prompt and delayed events.

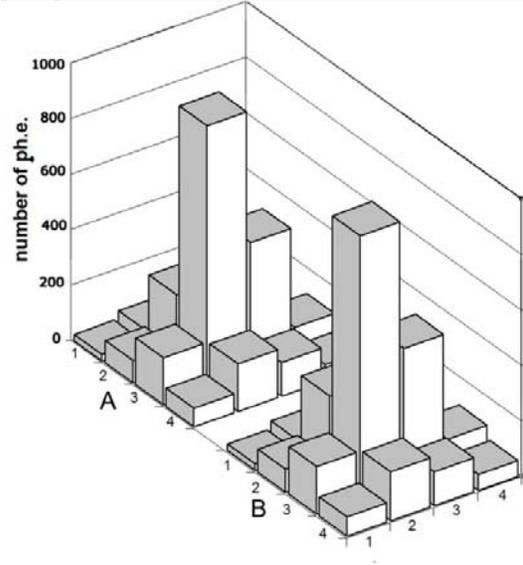

**Figure 2**. Simulated distribution of photoelectrons forming the hit pattern on the 16 × 16 PMT array for the readout of a scintillator bars array for the prompt event. A – PMTs at the "left" side of scintillation bars array, B – PMTs at the "right" side.

*3) Correlated background reduction:* Cosmogenic neutrons are produced in the atmosphere, in construction materials of the detector location, in the high-Z passive shielding of the detector, or within the detector itself. Since it is not possible to simulate the background caused by cosmogenic neutrons precisely because their flux is site dependent, instead the detector response to fast neutrons was simulated and the possibility to eliminate fast neutron events by applying topological cut by analysis of hit patterns was evaluated. Fast neutrons typically scatter on protons several times within a detector. Due to the relatively small (*n,p*) elastic scattering cross section, there is a considerable probability that in a segmented detector neutron will undergo scatters in many scintillation bars.

At each scattering neutron looses approximately half of its energy, therefore scintillation light caused by neutrons scattered elastically on the hydrogen in the scintillation bars is resulted in hit pattern, significantly different from PMT signal distribution of a antineutrino event. In contrast to antineutrino event fast neutron interaction characterized by spatially separated hits and high energy protons recoils. Additionally fast neutron events tend to localization close to detector edges.

For this purpose $2 \cdot 10^3$ fast neutrons in the energy range 1-50 MeV were simulated. The detector response to fast neutrons was derived from a data on neutrons slowing down history, stored in the output tally of MCNP.

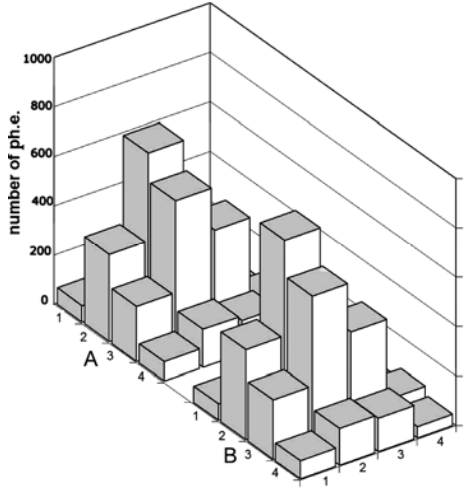

**Figure 3**. Simulated distribution of photoelectrons forming the hit pattern on the 16 × 16 PMT array for the readout of a scintillator bars array for the delayed neutron capture event. A and B as on Fig. 3.

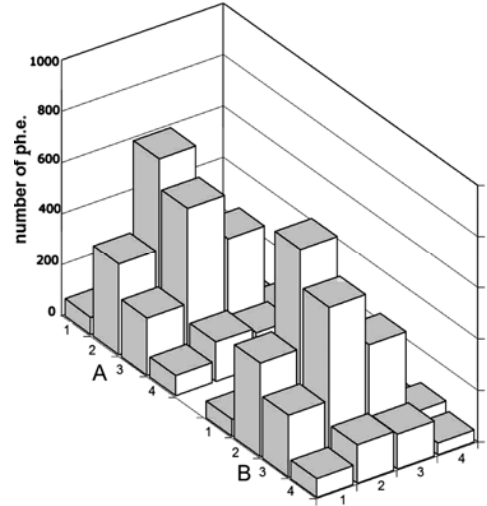

**Figure 4**. Simulated distribution of photoelectrons forming the hit pattern on the 16 × 16 PMT array for the readout of a scintillator bars array for the the fast neutron event with energy of $E$ = 15 MeV.

The energy deposits and positions of interaction of scattered protons in scintillation bars were transferred to ZEMAX transport code. Spherical light sources were attributed to positions of interaction with the number of rays quenched according to the energy dependence of light output of proton recoils in plastic scintillator given by

$$E_{ee} = 0.95 E_p - 8.0(1 - \exp(-0.1 E_p^{0.9})), \qquad (5)$$

where $E_p$ is the recoiling proton's energy, and $E_{ee}$ is the electron-equivalent light output in units of MeV [10,11].

The signal detected by the photomultipliers was used to form PMT signals distribution - hit pattern. The photoelectron distribution over the 16 × 16 PMT channels on both sides of scintillation bar array for fast neutron event in detector is shown on Fig. 4.

The hit patterns for prompt event of IBD are different from those for fast neutrons because for prompt event the center of gravity of hit pattern coincides with highest value of detected photoelectrons since annihilation gammas are emitted in opposite directions, so that there is symmetry in prompt event topology.

In a contrary, due to multiple scattering of fast neutrons the center of gravity of hit pattern is shifted from highest value of detected photoelectrons since more likely first scattered of proton carry out large amount of energy from initial fast neutron interaction with detector media.

Fast neutron event selection cut is made by applying $F$-test for equality of two variances to raw data [12]. This is performed by the statistical analysis of hit patterns simulated for fast neutrons and those stored in database corresponding to prompt events of IBD.

The plot of $F$-test values for simulated fast neutrons and prompt IBD events is presented on Fig. 5.

The $F$-test values for fast neutrons are grouped in the area of 0,1-0,8 and for IBD events they are nestled to 0,7-1 region.

Setting the $F$-test threshold to 0.75 will reject 82% of fast neutrons, but simultaneously rejecting 5% of IBD events. When setting $F$-test threshold to a value of 0.85, 95% of fast neutrons will be rejected, but 21% of IBD events will be lost. We decided to use a 0.75 $F$-test threshold with the hope to reject remaining fast neutron events but applying spatial analysis events distribution since fast neutrons in the energy range of interest trend to interact with detector close to its surface.

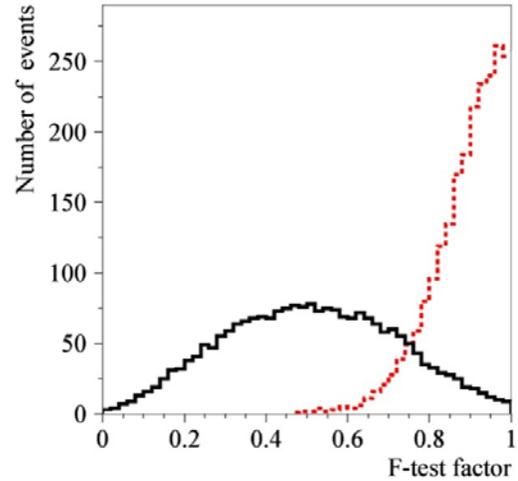

**Figure 5**. Fast neutrons and IBD events are discriminated by the use of $F$-test factor values. Dotted histogram corresponds to the $F$-test values for neutrons and continuous – for IBD events.

It should be noted, that detector response function to fast neutrons and actual efficiency of their subtraction should be determined experimentally by calibration with neutron sources.

*4) Uncorrelated background reduction:* Uncorrelated background is caused by accidental coincidence of $\beta^-/\gamma$-events with cosmogenic neutrons. To calculate the probability of random coincidence of $\beta^-/\gamma$-events with cosmogenic neutrons the prompt event caused by naturally occurring radioisotopes of $^{40}$K, $^{238}$U and $^{232}$Th that are present in PMT glass and detection media was simulated with the help of DECAY4 event generator and GEANT4 code [13]. Commercially are produced PMTs of standard type, low and ultra-low background. Two modifications of PMTs with different concentrations of radio-impurities [14] were considered. PMTs with standard borosilicate glass: $K < 60,000$ ppm, $Th < 1000$ ppb, $U < 1000$ ppb and PMT with low background glass: $K < 300$ ppm, $Th < 250$ ppb, $U < 100$ ppb. The radio-impurities of blocks made of plastic scintillator UPS-923A were taken from data of ZEPLIN-III experiment [15] and are as following: $U = (0.2 \pm 0.3) \times 10^{-9}$ ppb, $Th = (0.1 \pm 0.7) \times 10^{-9}$ ppb, $K = (0.2 \pm 0.6) \times 10^{-6}$ ppm.

The simulated background spectrum is presented on Fig. 6. The green line represent background from 32 PMTs with standard borosilicate glass, the red line background from 32 PMTs with low background glass and the black line is background from radioactive elements present in plastic scintillator. How it seen from the Fig. 7 $\beta^-/\gamma$-events coming from the $^{40}K$, $U$ and $Th$ chains do not extend beyond ~ 3 MeV. For a segmented detector one can calculate the rate of accidental events $R_{Acc}$ with

$$R_{Acc} = R_{prompt} \cdot R_{delayed} \cdot \Delta T \cdot \frac{V_r}{V_d}, \quad (6)$$

where $R_{prompt}$ is rate of events with energy fall inside energy window of prompt event, $R_{delayed}$ is the rate of events with energy fall inside energy window of the late event and $\Delta T$ is the time window, $V_r$ is volume of the cell accidental event is localized in, and $V_d$ is the volume of the detector. Taking into account results of simulation of detector spatial resolution the prompt event can be resolved within of $\approx (10\ cm)^3$ range and late event within of $\approx 8 \times (10\ cm)^3$. The single rate caused by spallation neutrons estimated at the level of 1 event per second. Then accidental coincidence rate of events from radioactive chains with cosmogenic neutrons in an energy range 2.0 – 8.0 MeV and in the time window of $\Delta T = 200\ \mu s$ will be less then $R_{Acc} \leq 10$ events/day for low background PMTs and $R_{Acc} \leq 100$ for the standard PMTs.

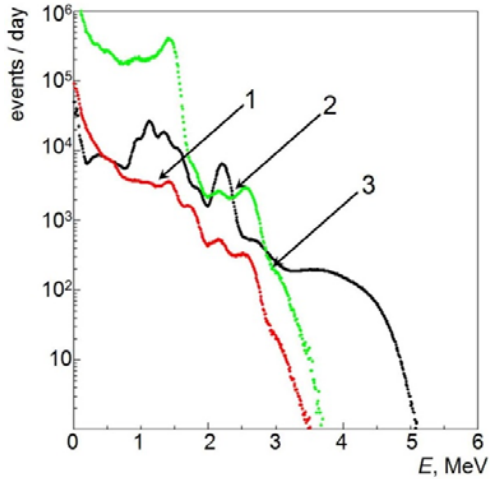

**Figure 6.** Simulated prompt energy spectrum for the accidental background caused by low background PMT – 1 (red), standard PMT – 2 (green) and internal radioactivity of plastic scintillator – 3 (black).

*5) Detector efficiency and performance:* In absence of oscillations, the number of antineutrino interactions in a detector target can be described as:

$$N_\nu = \frac{P_{th} \cdot N_p}{4\pi \cdot D^2 \langle E_f \rangle} \langle \sigma(E_\nu) \rangle \Phi_i(E_\nu), \quad (7)$$

where $N_p$ is the number of protons in the target, $D$ is the distance to the center of the reactor, and $P_{th}$ [GW] is the thermal power. $\langle E_f \rangle$ is the mean energy released per fission:

$$\langle E_f \rangle = \sum_k \alpha_k \langle E_f \rangle_k \quad (8)$$

where $\alpha_k$ is the fractional fission rate of the $k$-th isotope ($k$ = $^{235}U$, $^{239}Pu$, $^{238}U$, $^{241}Pu$). The mean cross section per fission $\langle \sigma_f \rangle$ is defined as

$$\sigma_f = \sum_k \alpha_k \sigma_{f\ k} = \sum_k \alpha_k \int_0^\infty dE_\nu S_k(E)\sigma_{IBD}(E_\nu) \quad (9)$$

where $S_k(E)$ is the reference spectrum of the $k$-th isotope and $\sigma_{IBD}$ is the inverse beta decay cross section. The three variables $P_{th}$, $\langle E_f \rangle$, and $\langle \sigma_f \rangle$ are time dependent, with $\langle E_f \rangle$ and $\langle \sigma_f \rangle$ depending on the evolution of the fuel composition in the reactor and $P_{th}$ depending on the operation of the reactor.

About 6400 interactions/day can be expected in a 1 ton plastic scintillator target placed 12 m from a 1.4 GW$_{th}$ power reactor VVER-440.

With the purpose of IBD reactions selection from a background which on a few orders of magnitude exceeds the signal, the selection cuts are applied to the simulation outputs. Prompt and delayed events were selected using the sum of the energy deposits of all scintillation bars. The energy window for the prompt event is chosen in the range of 2,0–8 MeV,

TABLE I
EFFICIENCY OF ANTINEUTRINO DETECTION.

| Selection cut | Efficiency |
|---|---|
| Prompt signal energy window ($2 < E_\nu < 8$ MeV) | 88% |
| Prompt event topological cut | 94% |
| Delayed signal energy window ($3 < E_n < 8$ MeV) | 95% |
| Probability of neutron capture by the Gd nucleus | 75% |
| Time window ($2\ \mu s < \Delta t < 200\ \mu s$) | 95% |
| Total efficiency | 56% |

which is based on simulation results of expected accidental coincidence rate.

The energy distribution of the IBD positrons is directly related to the antineutrino energy distribution. But only a few prompt events with $E_{total}$ beyond 8 MeV contributes to total antineutrino signal. The delayed energy window was chosen in the range of 3−9 MeV. Since some cascading gamma rays from neutron capture by Gd may escape detector active volume, the lowest threshold was set to be higher than 3 MeV with purpose to reject background gamma rays but low enough to retain as many cascading gamma rays as possible. The effects of the selection cuts on antineutrino detection efficiency are listed in Table I.

For this detector geometry and applied cuts antineutrino detection efficiency is estimated at the level of 55%, and expected antineutrino counting rate is $N_\nu \approx 3500$ events/day.

*B. Fuel composition monitoring*

*1) Calculation of reactor antineutrino spectra:* In order to test the sensitivity of proposed detector to measure fuel composition evolution during fuel cycle the calculation of reactor antineutrino spectra was performed. The total reactor antineutrino energy spectrum is a sum of weighted individual antineutrino spectra from the main uranium and plutonium isotopes with their fission rates:

$$S(E) = \sum_k \alpha_k S_k(E), \quad (10)$$

where $\alpha_k$ is the fractional fission rate of the $k$-th isotope defined in Eq. (9). The $S_k(E)$ represents the total $\nu$ spectrum emitted by a reactor per fission by the $k$-th isotope and is calculated by summation method by adding all fission product $\beta/\nu$ spectra weighted by their activity $A_f$ of the fission product,

$$S_k(E) = \sum_{f=1}^{N_f} A_f S_f(E), \qquad (11)$$

where spectrum $S_f(E)$ of each fission product is a sum of $N_b$ β-branches connecting the ground state (or an isomeric state) of the parent nucleus to different excited levels of the daughter nucleus:

$$S_f(E) = \sum_{b=1}^{N_b} BR_f^b S_f^b(Z_f, A_f, E_{0f}^b, E), \qquad (12)$$

where $BR_f^b$ and $E_{0f}^b$, being the branching ratio and the endpoint energy of the $b$ branch of the $f$ fission product, respectively, and $Z_f$ and $A_f$ being the charge and atomic number of the parent nucleus. We have used nuclear data from ENSDF, ENDF/B.VII.1 and nuclear data libraries included in ORIGEN-S from Scale program (decay data, branching ratios, etc.) [16, 17, 18,19]. The calculated antineutrino energy spectra of $^{235}$U, $^{238}$U, $^{239}$Pu and $^{241}$Pu folded with cross-section for inverse beta decay are presented on Fig. 7.

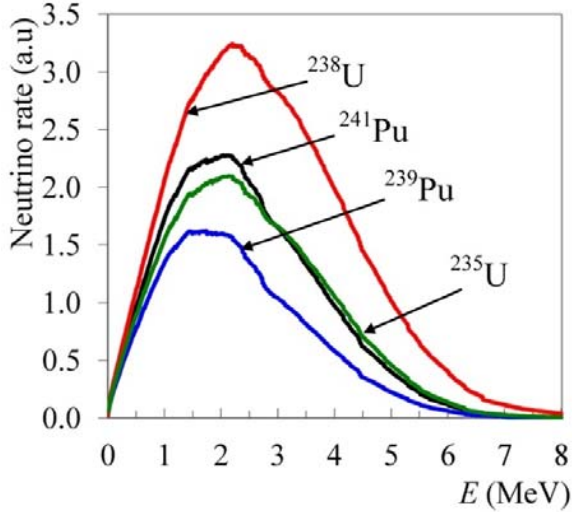

**Figure 7.** Calculated positron spectra: 1 – $^{235}$U (green line), 2 – $^{239}$Pu (blue line), 3 – $^{238}$U (red line) and 4 – $^{241}$Pu (black line).

The sum spectrum of this four main isotopes calculated for the middle of reactor fuel burning cycle describe well main features of antineutrino spectrum measured by near detector of RENO collaboration [20] as one can see on Fig. 8.

To test the possibility of proposed detector to perform independent measurement of fuel composition evolution during reactor fuel cycle we have simulated and analyzed total antineutrino spectrum emitting by nuclear reactor at different periods of fuel burning cycle.

*2) Energy resolution:* Calculated light collection efficiency of the proposed detector ~19% combined with high light output of $10^4$ photons per 1 MeV of electron equivalent energy of incident particle and 25% quantum efficiency of standard PMTs will allow obtaining a signal efficiency of 475 photoelectrons/MeV. Based on this value the reconstruction gives a full width half maximum (FWHM) positron's energy resolution of $\delta E/E \sim 11\%/\sqrt{E(MeV)}$. Since not all possible effects that may worsen detector energy resolution included in simulation we used 15% FWHM at 1 MeV energy resolution for analyzing detector sensitivity for fuel evolution. The antineutrino energy spectrum was generated as the sum of four main isotopes $^{235}$U, $^{238}$U, $^{239}$Pu and $^{241}$Pu that contribute to detected varied in time according to data of Rivno NPP operator

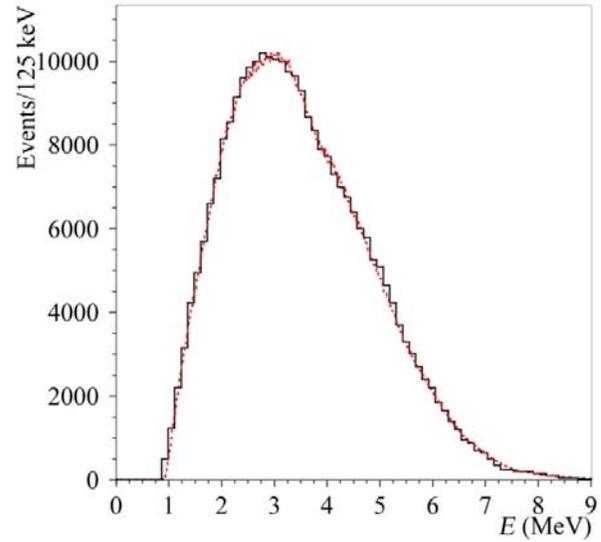

**Figure 8.** Spectra of IBD prompt signal: calculated - red dashed line, measured by near detector of RENO collaboration [20] - black continuous line.

presented in antineutrino rate with the each isotope fraction Table II [30]. Then detected antineutrino spectra were obtained by folding the calculated spectrum with detector respond

TABLE II
A FUEL COMPOSITION IN RELATIVE NUMBER OF
FISSIONS FOR MAIN FISSION ISOTOPES
(from ref. [30]).

| Date | Relative number of fissions | | | |
|---|---|---|---|---|
| | $^{235}$U | $^{239}$Pu | $^{238}$U | $^{241}$Pu |
| 23/03/1988 | 54,6 | 32,3 | 7,6 | 5,5 |
| 02/05/1988 | 52,6 | 33,7 | 7,7 | 6 |
| 02/07/1988 | 49,6 | 35,6 | 7,8 | 7 |
| Fuel reloading | | | | |
| 01/08/1988 | 71,2 | 19,2 | 7,1 | 2,5 |
| 11/09/1988 | 67,4 | 22,4 | 7,3 | 2,9 |
| 23/10/1988 | 64,3 | 25 | 7,3 | 3,4 |
| 08/12/1988 | 61,5 | 27,2 | 7,4 | 3,9 |

function, thus distortions induced by the detector energy resolution where taken into account.

On Fig. 9 we have reproduced the antineutrino energy spectra accumulated over 5 days before reactor stop for refueling and 5 days after reactor's operating cycle start for a detector of a 1 m$^3$ target volume, which is assumed to be installed 12 m away from a 1.4 GW$_{th}$ VVER-440 reactor.

The 5 days integrated antineutrino energy spectra corresponding to fuel composition from Table II corrected by the detector efficiency and energy resolution were analyzed by means of different statistical methods. It was found that $\chi^2$ and Kolmogorov-Smirnov tests are less sensitive to total antineutrino spectra changing during the fuel cycle in contrast to $F$-test or Mann − Whitney $U$-test as it can be seen from Fig. 10.

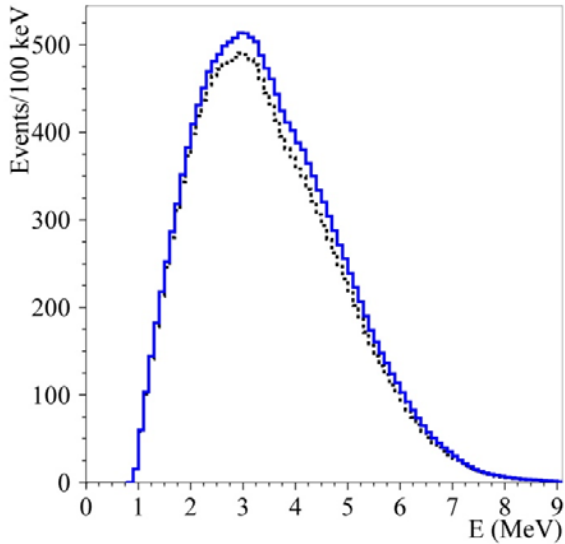

**Figure 9.** Detected antineutrino spectra integrated over 5 days at the beginning of fuel burning cycle (blue continuous line) and at the end of a cycle (black dashed line).

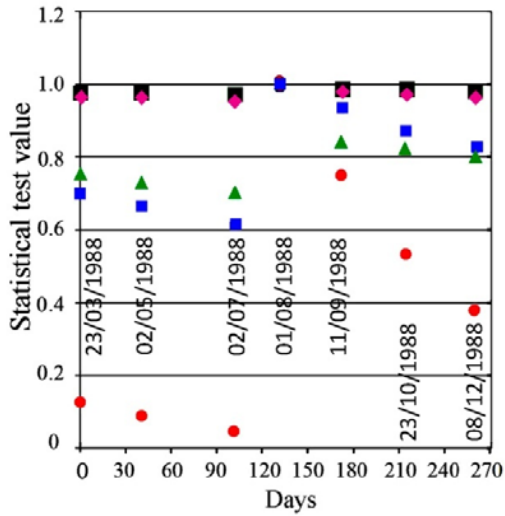

**Figure 10**. Statistical tests values for analysis of detected antineutrino spectra during reactor burning cycle for 15%/MeV detector energy resolution: $\chi^2$ – test – black squares, Kolmogorov-Smitnov test – magenta diamonds, $F$-test – blue squares, $U$-test – green triangles. $F$-test values for ideal detector – red cycles.

One can see that spectra shape analysis with the help of $F$-test, when no energy resolution has been taken into account, shows considerable variation of shapes during fuel cycle. When energy resolution have been considered, the statistical analysis shows less difference in spectra shapes but $F$-test and $U$-test are still applicable to determine if datasets are different according changing of fuel composition during fuel cycle. One can see that real energy resolution of detector of 15%/MeV still allow to control fuel composition evolution of reactor core during fuel cycle.

The effect of ~12% fractional change in the antineutrino energy spectrum as a function of energy as simulated at the beginning and end of a representative 270 days of VVER-440 fuel burning cycle can be detected with expected detector energy resolution (see Fig. 11).

IV. CONCLUSIONS

The novel design of a segmented antineutrino detector is developed that utilizes existing technologies and its capabilities have been demonstrated through extensive Monte Carlo simulations. The applied block-design scheme widely used in Positron Emission Tomography scanners allow drastically reduce the number of photo-detectors and output readouts comparing to one-to-one coupling between scintillation bar and photo-detectors that makes the data acquisition system much simpler and cheaper. All this features allow building detector in a cost effective way.

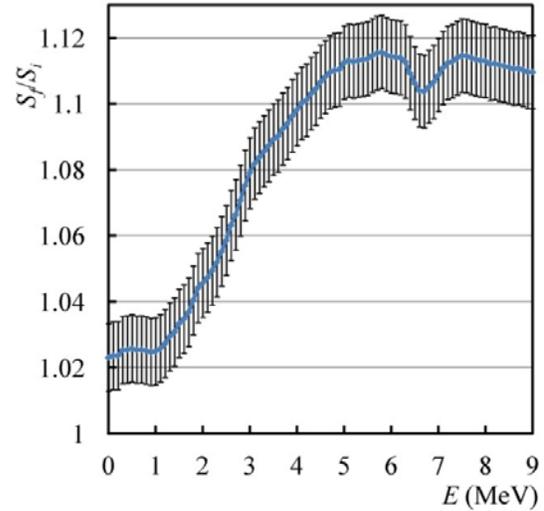

**Figure 11.** Antineutrino spectra shape ratio at the beginning $S_i$ and the end $S_f$ of fuel burning cycle.

The expected energy resolution of 11-15%/Mev and detection efficiency of ~55% will allow independent monitoring of evolution of reactor fuel isotope composition during the reactor operation. The detector design provides easy construction and installation and safety of it exploitation.

Placing the detector in reactor building of power units 2 of Rivne NPP (VVER-440) at the distance of 12 meters from a reactor core the antineutrino count rate of ~3500 events/day is expected.